\documentclass[a4paper,11pt]{article}
\usepackage{pos}
\usepackage{orcidlink}
\usepackage{soul}

\title{Configurational Thermometer for Lattice Gauge Theories}
\ShortTitle{Configurational Thermometer for Lattice Gauge Theories}

\author*[a]{Vamika Longia~\orcidlink{0009-0004-2542-2805}}
\affiliation[a]{Department of Physical Sciences, Indian Institute of Science Education and Research - Mohali, Knowledge City, Sector 81, SAS Nagar, Punjab 140306, India}
\emailAdd{vamika.longia@gmail.com}

\author[b]{Navdeep Singh Dhindsa~\orcidlink{0000-0002-3133-6979}}
\affiliation[b]{Department of Theoretical Physics, Tata Institute of Fundamental Research,\\Homi Bhabha Road, Mumbai 400005, India}
\emailAdd{navdeep@theory.tifr.res.in}

\author[c]{Anosh Joseph~\orcidlink{0000-0003-4288-8207}}
\affiliation[c]{National Institute for Theoretical and Computational Sciences, \\ School of Physics, and Mandelstam Institute for Theoretical Physics,\\ University of the Witwatersrand, Johannesburg, Wits 2050, South Africa}
\emailAdd{anosh.joseph@wits.ac.za}

\FullConference{The 42nd International Symposium on Lattice Field Theory (LATTICE2025)\\
2-8 November 2025\\
Tata Institute of Fundamental Research, Mumbai, India\\}

\abstract{
We propose a diagnostic tool, a temperature estimator, for lattice gauge theory simulations. The estimator is obtained from the gradient and the Hessian of the Euclidean lattice action. It is gauge invariant, configuration-based, and independent of momentum-space information. These features enable direct checks of thermodynamic consistency in Monte Carlo simulations. We apply this tool to compact U(1) lattice gauge theories in one, two, and four dimensions. The results confirm the proposed estimator's ability to reproduce the input temperatures across different lattice ensembles. The estimator is sensitive to sampling inefficiencies and algorithmic artifacts, making it a useful diagnostic for large-scale simulations.
}

\begin{document}
\maketitle

\section{Introduction}
\label{sec:intro}

A basic consistency requirement of numerical simulations of statistical and quantum field theories is the faithful sampling of configurations corresponding to the chosen parameter set. However, in practice, effects such as metastability or false vacua, critical slowing down, long autocorrelation times, or subtle implementation errors can lead to effective sampling that differs from the intended regime. These issues are likewise relevant in lattice gauge theories, where the dynamics of all lattice fields are captured through stochastic sampling. Hence, it is important to monitor this directly, motivating us to introduce an observable that extracts thermodynamic information from the configurations themselves, without relying on auxiliary variables or indirect equilibrium assumptions.

A geometric approach to thermodynamic temperature, originally developed by Rugh \cite{Rugh:1997,Rugh:1998}, gives an alternative framework in which temperature is expressed in terms of phase-space geometry rather than energy fluctuations or momenta. Building on this idea, we introduce a configurational temperature estimator for lattice field theories that depends only on the gradient and Hessian of the Euclidean lattice action \cite{Dhindsa:2025xfv}. The resulting estimator is purely configurational, gauge invariant, and can be evaluated directly on field configurations generated in standard Monte Carlo simulations.

When applied to lattice gauge theories, the estimator can provide a microscopic diagnostic of thermodynamic consistency that does not rely on a specific ensemble generation method. It allows one to directly compare the effective temperature extracted from the sampled configurations with the input temperature used in the simulation. Systematic deviations in the measured temperature can therefore indicate sampling bias or implementation errors and serve as a useful consistency check.

In this work, we benchmark the configurational temperature estimator using compact U(1) lattice gauge theories in one, two, and four Euclidean dimensions. These theories provide controlled settings ranging from exactly solvable models to systems exhibiting nontrivial phase structure. In all the cases, we find that the estimator reliably reproduces the target temperature over a wide range of couplings and lattice sizes. We also demonstrate that the estimator can clearly identify algorithmic issues that are not readily detectable from standard observables. Our results establish the configurational temperature as a practical and robust diagnostic tool for assessing stability and consistency in lattice gauge theory simulations, with natural extensions to more general gauge groups and numerical frameworks.

\section{Derivation of the Configurational Temperature Estimator}
\label{sec:ensemble}

We briefly review the geometric formulation of temperature in the microcanonical ensemble and introduce the configurational temperature estimator, which serves as the basis for our analysis.

Consider a classical system with phase space coordinates
\begin{equation}
\vec{\Gamma} \equiv (q_1, q_2, \dots, q_N; p_1, p_2, \dots, p_N),
\end{equation}
where $q_i$ and $p_i$ denote the position and momentum of the $i$-th degree of freedom. The microcanonical ensemble at fixed energy $E$ consists of all phase space points lying within a thin energy shell $\mu C(E)$ satisfying the condition,
\begin{equation}
E - \frac{1}{2}\delta E \leq H(\vec{\Gamma}) \leq E + \frac{1}{2}\delta E,
\end{equation}
with $\delta E \ll E$, typically scaling as $\delta E/E = \mathcal{O}(1/\sqrt{N})$. Assuming equal a priori probabilities, the entropy of the ensemble is defined as
\begin{equation}
S(E) = k_B \ln \int_{\mu C(E)} d\vec{\Gamma}.
\end{equation}

The geometric formulation of temperature is obtained by considering an infinitesimal deformation of the energy shell in phase space. Restricting attention to purely configurational variations and keeping the momenta fixed, we define a shift of phase space points
\begin{equation}
\label{eq:translation}
\vec{\Gamma} \rightarrow \vec{\Gamma}' = \vec{\Gamma} + \Delta E \, \vec{n}(\vec{\Gamma}),
\end{equation}
where the vector field $\vec{n}(\vec{\Gamma})$ is chosen to generate a uniform increase in energy across the ensemble at leading order in $\Delta E$. A convenient choice is
\begin{equation}
\vec{n}(\vec{\Gamma}) =
\frac{\vec{\nabla}_{\vec{q}} H(\vec{\Gamma})}
{\vec{\nabla}_{\vec{q}} H(\vec{\Gamma}) \cdot \vec{\nabla}_{\vec{q}} H(\vec{\Gamma})},
\end{equation}
where $\vec{\nabla}_{\vec{q}}$ denotes the configurational gradient,
\begin{equation}
\vec{\nabla}_{\vec{q}} = \left( \frac{\partial}{\partial q_1}, \dots, \frac{\partial}{\partial q_N} \right).
\end{equation}
With this choice, the displaced configuration satisfies
\begin{equation}
\vec{\Gamma}' = \vec{\Gamma} + \Delta E
\frac{\vec{\nabla}_{\vec{q}} H(\vec{\Gamma})}
{\vec{\nabla}_{\vec{q}} H(\vec{\Gamma}) \cdot \vec{\nabla}_{\vec{q}} H(\vec{\Gamma})}.
\end{equation}
Under this transformation, the entropy of the displaced ensemble becomes
\begin{equation}
S(E + \Delta E) = k_B \ln \int_{\mu C(E + \Delta E)} d\vec{\Gamma}'.
\end{equation}
Introducing the Jacobian of the transformation, this may be written as
\begin{equation}
S(E + \Delta E) = k_B \ln \int_{\mu C(E)}
\left| \frac{\partial \vec{\Gamma}'}{\partial \vec{\Gamma}} \right|
d\vec{\Gamma}.
\end{equation}
Substituting Eq.~\eqref{eq:translation} and expanding to leading order in $\Delta E$, one finds
\begin{equation}
S(E + \Delta E) = k_B \ln \int_{\mu C(E)}
\left( 1 + \Delta E \, \vec{\nabla}_{\vec{q}} \cdot \vec{n}(\vec{\Gamma}) \right)
d\vec{\Gamma}.
\end{equation}

The change in entropy is therefore
\begin{equation}
\begin{aligned}
\Delta S &= S(E + \Delta E) - S(E) \\
&= k_B \ln \left( 1 + \Delta E
\left\langle \vec{\nabla}_{\vec{q}} \cdot \vec{n}(\vec{\Gamma}) \right\rangle \right) \\
&\approx k_B \Delta E
\left\langle \vec{\nabla}_{\vec{q}} \cdot \vec{n}(\vec{\Gamma}) \right\rangle,
\end{aligned}
\label{eq:entropy_change}
\end{equation}
where $\langle \cdots \rangle$ denotes the microcanonical average and we have used $\ln(1+x)\simeq x$ for small $x$.

From the first law of thermodynamics, $dE = T dS - p dV$, the inverse temperature at fixed volume is defined as
\begin{equation}
\label{eq:temperature}
\frac{1}{T} = \left( \frac{\partial S}{\partial E} \right)_V.
\end{equation}
Taking the limit $\Delta E \to 0$ in Eq.~\eqref{eq:entropy_change}, we obtain
\begin{equation}
\label{eq:partialS}
\frac{\partial S}{\partial E}
= k_B \left\langle \vec{\nabla}_{\vec{q}} \cdot \vec{n}(\vec{\Gamma}) \right\rangle,
\end{equation}
and hence
\begin{equation}
\frac{1}{k_B T}
= \left\langle
\vec{\nabla}_{\vec{q}} \cdot
\left(
\frac{\vec{\nabla}_{\vec{q}} H(\vec{\Gamma})}
{\vec{\nabla}_{\vec{q}} H(\vec{\Gamma}) \cdot \vec{\nabla}_{\vec{q}} H(\vec{\Gamma})}
\right)
\right\rangle.
\end{equation}

For purely configurational systems, or when the Hamiltonian is dominated by a potential term $H(\vec{q}) = \Phi(\vec{q})$, this expression reduces to
\begin{equation}
\frac{1}{k_B T}
= \left\langle
\vec{\nabla}_{\vec{q}} \cdot
\left(
\frac{\vec{\nabla}_{\vec{q}} \Phi(\vec{q})}
{\vec{\nabla}_{\vec{q}} \Phi(\vec{q}) \cdot \vec{\nabla}_{\vec{q}} \Phi(\vec{q})}
\right)
\right\rangle
- \mathcal{O}\!\left(\frac{1}{N}\right).
\end{equation}
For finite systems, replacing the full microcanonical average by a configurational one introduces corrections suppressed by system size, which vanish in the thermodynamic limit. 

Introducing the gradient and Hessian of the potential,
\begin{equation}
\vec{g} = \vec{\nabla}_{\vec{q}} \Phi(\vec{q}), \qquad \vec{g} \in \mathbb{R}^N,
\end{equation}
\begin{equation}
\mathbb{H} = \vec{\nabla}_{\vec{q}} \vec{\nabla}_{\vec{q}}^{\,T} \Phi(\vec{q}),
\qquad \mathbb{H} \in \mathbb{R}^{N \times N},
\end{equation}
the configurational temperature estimator is defined as
\begin{equation}
\beta_M \equiv
\left\langle
\vec{\nabla}_{\vec{q}} \cdot
\left(
\frac{\vec{g}}{\vec{g} \cdot \vec{g}}
\right)
\right\rangle = \left \langle \frac{{\rm Tr} ({\mathbb H})}{| \vec{g} |^2} - 2 \frac{\vec{g}^T {\mathbb H} \vec{g}}{|\vec{g}|^4} \right \rangle.
\end{equation}

Although the estimator is derived using geometric arguments in the microcanonical ensemble, we can justify its application in canonical Monte Carlo simulations under standard assumptions of ensemble equivalence in the thermodynamic limit. For systems with sufficiently many degrees of freedom and short-range interactions, microcanonical and canonical ensembles yield identical expectation values up to finite-size corrections. Since $\beta_M$ depends only on local configurational derivatives of the action, it probes geometric properties of the energy landscape rather than the global energy fluctuations. As a result, when configurations are sampled from a canonical distribution at inverse temperature $\beta$, the estimator converges to the same thermodynamic temperature, provided the system is well thermalized and free from sampling bias. Deviations of $\beta_M$ from the input $\beta$ therefore signal violations of these assumptions, such as incomplete equilibration, metastability, or algorithmic errors, rather than a failure of ensemble equivalence itself.

\section{U(1) Lattice Gauge Theories}
\label{sec:benchmark-result}

To demonstrate the applicability of the configurational temperature estimator, we benchmark it using compact U(1) lattice gauge theory in various Euclidean spacetime dimensions. 
This class of models provides a controlled sequence of test cases with progressively richer dynamics and phase structure. 
In one Euclidean dimension, pure U(1) lattice gauge theory has no local dynamical degrees of freedom and consequently exhibits no thermodynamic phase transition in the usual sense. 
In two dimensions, compact U(1) lattice gauge theory is exactly solvable and remains confining for all values of the coupling, with no phase transition separating distinct phases. 
In three dimensions, the lattice theory likewise exhibits confinement for all couplings, driven by monopole condensation \cite{Caselle:2025bgu}. 
By contrast, in four dimensions the theory possesses two distinct phases: a confining phase at strong coupling and a Coulomb phase at weak coupling, characterized by a massless photon and a broken global U(1)symmetry \cite{Vettorazzo:2004cr}. 
Taken together, these features make compact U(1) lattice gauge theory a well-suited testing ground for validating the estimator both in analytically controlled regimes and in the presence of nontrivial lattice phase structure.

The gauge degrees of freedom are defined on the links of a hypercubic lattice and take values in the U(1) group,
\begin{equation}
U_{\mu}(n) = e^{i \theta_{\mu}(n)},
\end{equation}
with $\theta_{\mu}(n) \in (-\pi,\pi]$, where $\mu = 1, \dots , D$ labels the Euclidean spacetime directions.

\subsection{One-dimensional theory}

We begin with the one-dimensional compact U(1) lattice gauge theory, which serves as a simple and analytically tractable benchmark. In this formulation, the gauge degrees of freedom are compact variables $U_n$ residing on the links connecting neighboring lattice sites $n$ and $n+1$. Each link variable is an element of the U(1) group and may be parameterized as
\begin{equation}
U_n = e^{i\theta_n}, \qquad \theta_n \in (-\pi, \pi].
\end{equation}

With this parametrization, the Wilson gauge action reduces to a nearest-neighbor interaction along the lattice direction,
\begin{equation}
S = \beta \sum_{n = 0}^{N_\tau - 1}
\left[ 1 - \cos(\theta_n - \theta_{n+1}) \right],
\end{equation}
where $N_\tau$ denotes the total number of lattice sites.

\begin{figure*}[t]
\centering
\includegraphics[width=0.32\textwidth]{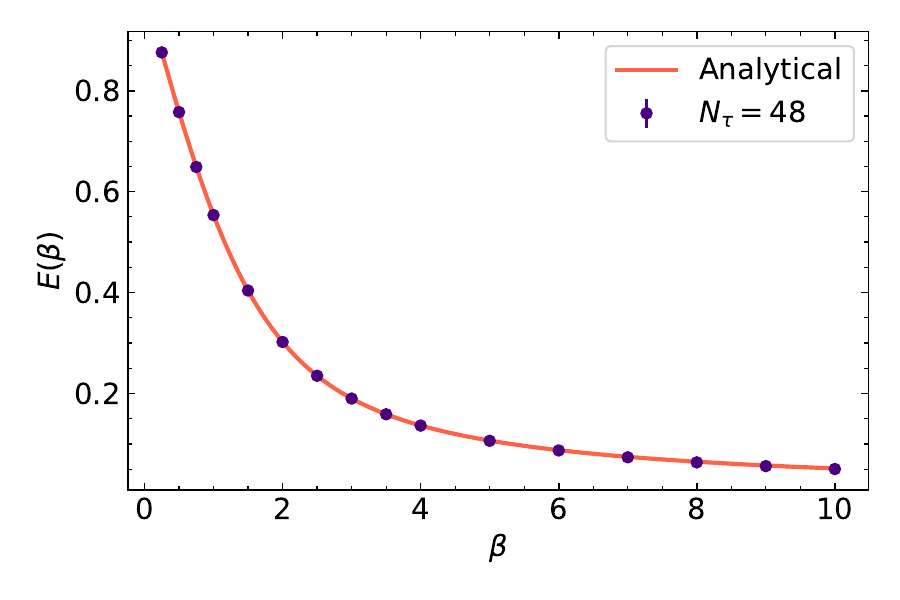}
\includegraphics[width=0.32\textwidth]{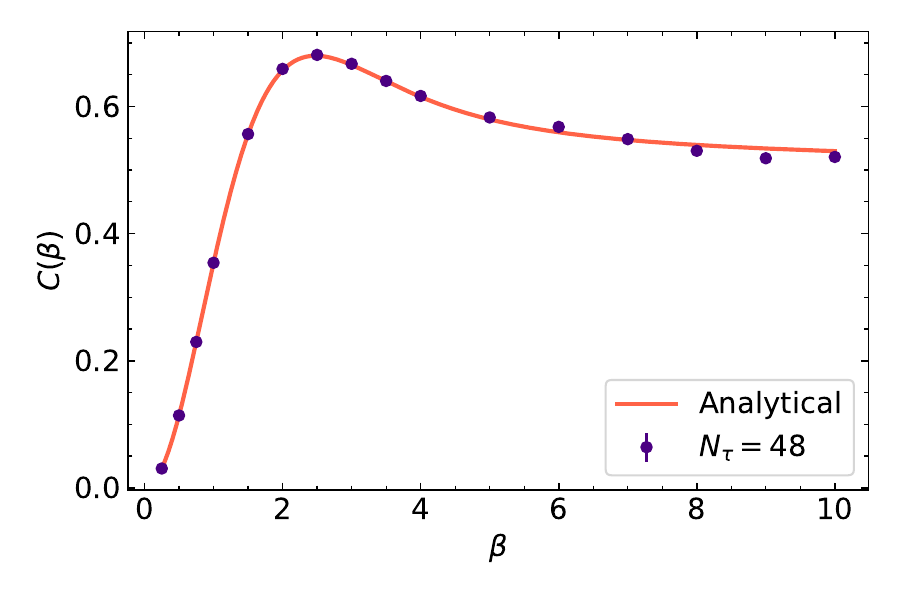}
\includegraphics[width=0.32\textwidth]{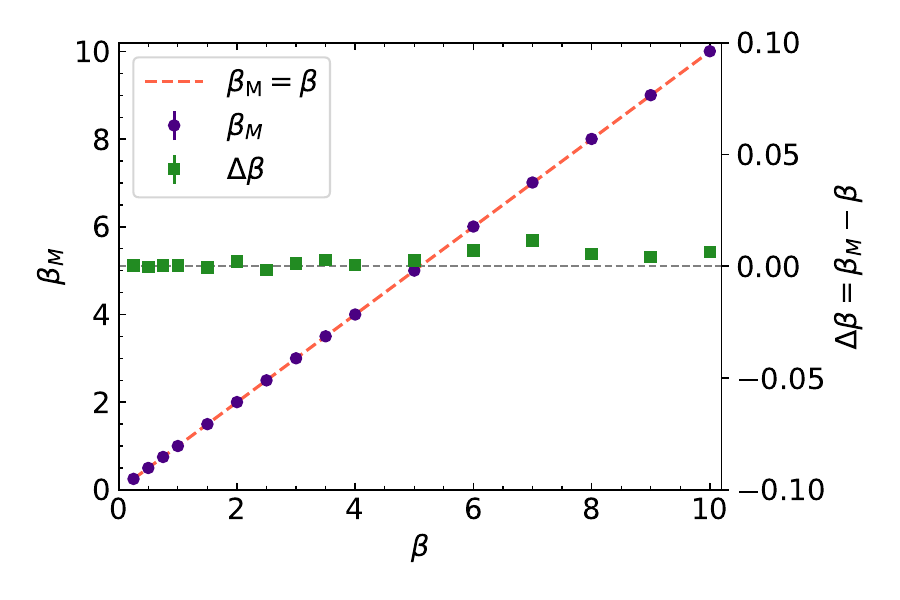}
\caption{Thermodynamic observables for the one-dimensional U(1) lattice theory on a 48-site lattice. The energy $E(\beta)$ (left), specific heat $C(\beta)$ (middle), and the measured inverse temperature $\beta_M$ (right) are plotted against the input $\beta$.}
\label{fig:1d_u1_l48}
\end{figure*}

We generate gauge field configurations using the Hybrid Monte Carlo algorithm and compute thermodynamic observables including the average energy, the specific heat, and the configurational temperature estimator $\beta_M$. Results for a lattice with $N_\tau = 48$ sites are shown in Fig.~\ref{fig:1d_u1_l48}. As illustrated in the left and middle panels, the numerically measured energy $E(\beta)$ and specific heat $C(\beta)$ agree with the exact analytical expressions. The right panel shows the estimator $\beta_M$, which accurately reproduces the input $\beta$ over the entire range studied, thereby validating the estimator in this exactly solvable setting.

\subsection{Two-dimensional theory}

We next consider compact U(1) lattice gauge theory in two Euclidean dimensions. The Wilson action takes the form
\begin{equation}
S = \beta \sum_{(x, y)} \left[ 1 - \cos\theta_p(x, y)\right],
\end{equation}
where $\theta_p(x, y)$ denotes the plaquette angle at lattice site $(x, y)$, defined as the oriented sum of link variables around the elementary square,
\begin{equation}
\theta_p(x, y) = \theta_x(x, y) + \theta_y(x + 1, y)
- \theta_x(x, y + 1) - \theta_y(x, y),
\end{equation}
with $\theta_{x/y}(x, y) \in (-\pi, \pi]$.

\begin{figure*}[t]
\includegraphics[width=0.3\textwidth]{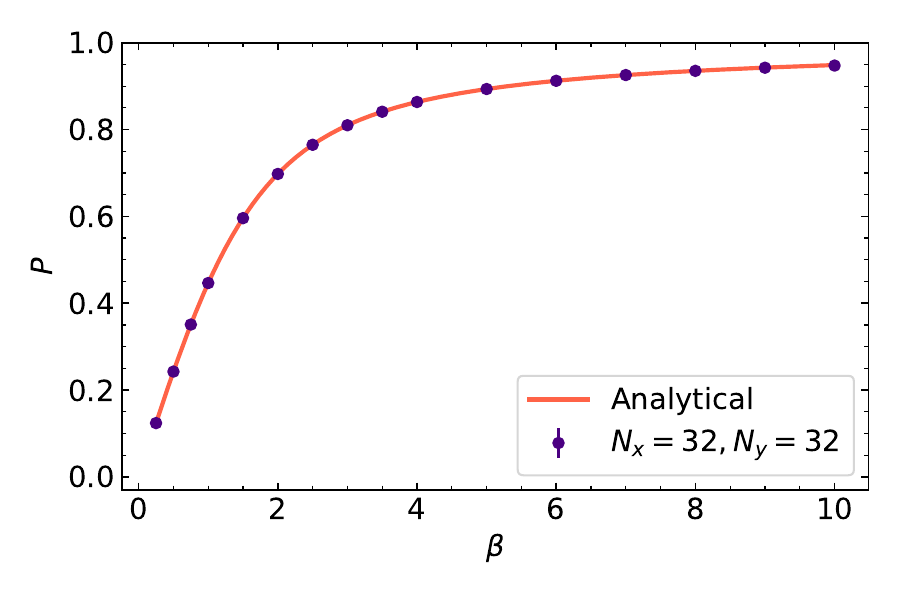}
\includegraphics[width=0.3\textwidth]{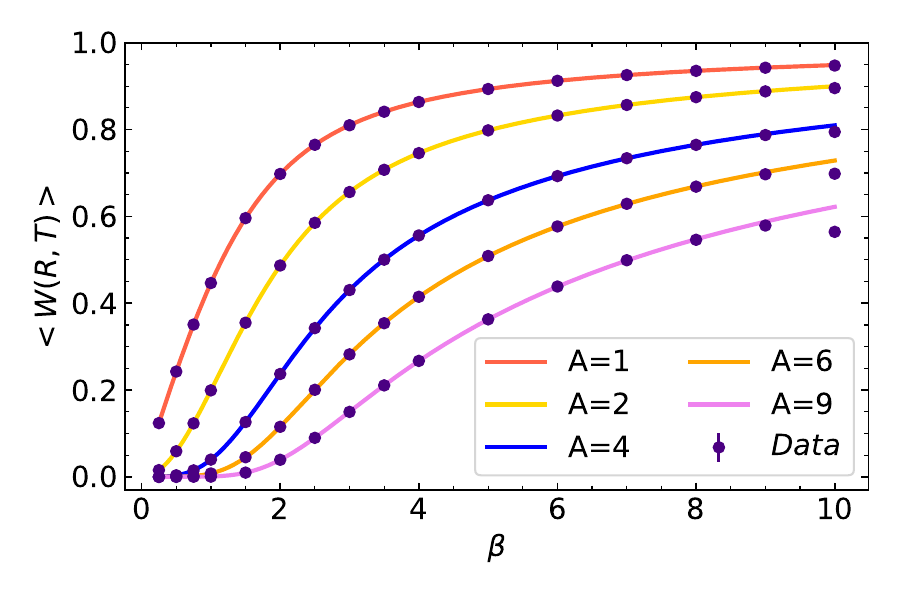}
\includegraphics[width=0.3\textwidth]{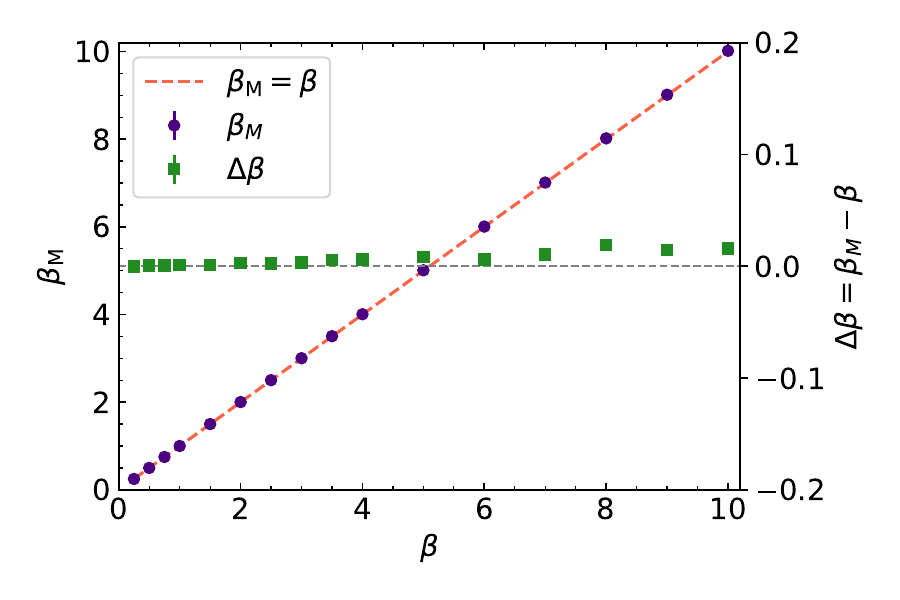}
\caption{\label{fig:plaquette_beta_2d}
Thermodynamic observables for the two-dimensional U(1) lattice gauge theory on a $32 \times 32$ lattice. The expectation value of the plaquette (left), Wilson loops of sizes $A = R \times T$ (middle), and the estimated inverse temperature $\beta_M$ (right) are shown as functions of the input $\beta$.}
\end{figure*}

Since the two-dimensional theory is exactly solvable, it provides another stringent test of the estimator. On a $32 \times 32$ lattice, we compute the plaquette expectation value $P$, and Wilson loops $W(R, T)$ with spatial and temporal extents $R$ and $T$. The measured observables are compared with their analytical counterparts, while the configurational temperature $\beta_M$ is directly compared with the input coupling $\beta$. As shown in Fig.~\ref{fig:plaquette_beta_2d}, the estimator reliably tracks the input temperature across the full coupling range, in parallel with the excellent agreement observed for $P$ and $\langle W(R,T)\rangle$.

\subsection{Four-dimensional theory}

Having validated the estimator in one and two dimensions, we finally turn to the four-dimensional compact U(1) lattice gauge theory, which exhibits a nontrivial phase transition on the lattice. The Wilson action in four Euclidean dimensions is
\begin{equation}
S = \beta \sum_x \sum_{\mu < \nu}
\left[ 1 - \cos\theta_{\mu\nu}(x) \right],
\label{eq:u1_4d_action}
\end{equation}
where $x$ labels lattice sites and $\mu, \nu \in \{0,1,2,3\}$ denote spacetime directions. The plaquette angle $\theta_{\mu\nu}(x)$ is given by
\begin{equation}
\theta_{\mu \nu}(x) =
\theta_\mu(x) + \theta_\nu(x + \hat\mu)
- \theta_\mu(x + \hat\nu) - \theta_\nu(x),
\end{equation}
with $\theta_\mu(x) \in (-\pi, \pi]$.

\begin{figure*}[t]
\centering
\includegraphics[width=0.32\textwidth]{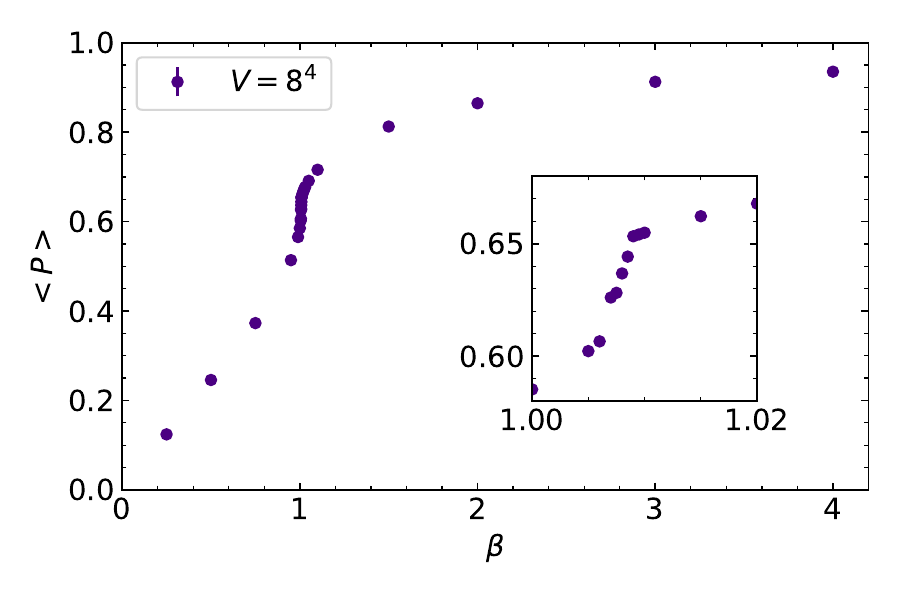}
\includegraphics[width=0.32\textwidth]{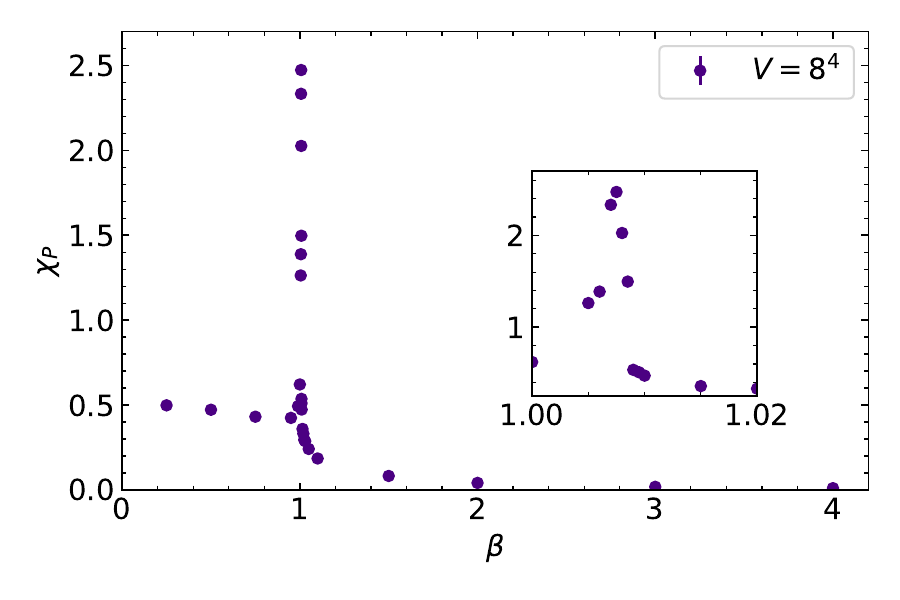}
\includegraphics[width=0.32\textwidth]{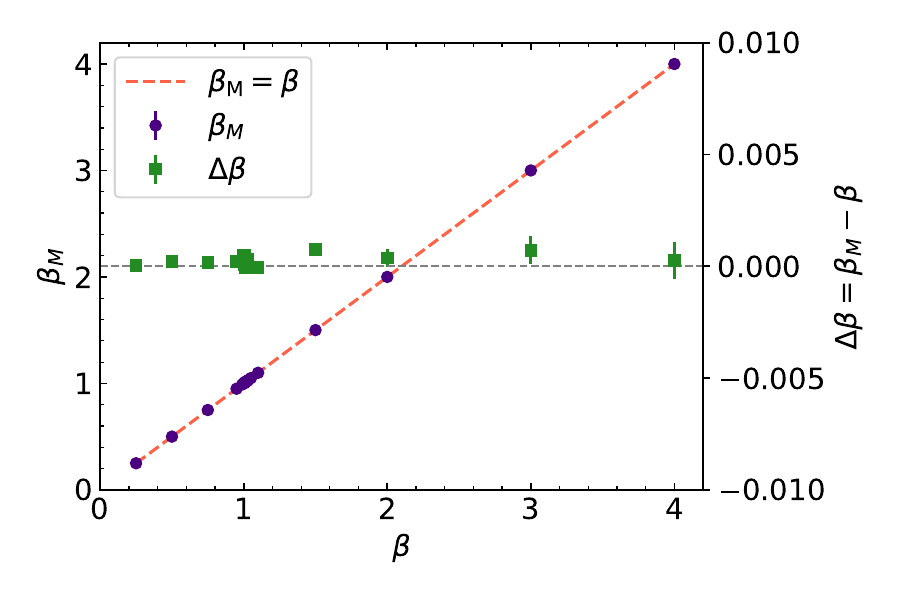}
\caption{Thermodynamic observables for the four-dimensional U(1) lattice theory on an $8^4$ lattice. The plaquette $\langle P \rangle$ (left), the plaquette susceptibility $\chi_P$ (middle), and the estimated inverse temperature $\beta_M$ (right) are plotted against the input $\beta$.}
\label{fig:4d_u1_8}
\end{figure*}

On an $8^4$ lattice, we compute the plaquette expectation value $\langle P \rangle$, the plaquette susceptibility $\chi_P$, and the configurational temperature estimator $\beta_M$ as functions of the input $\beta$. The results are shown in Fig.~\ref{fig:4d_u1_8}. The standard observables exhibit the expected behavior near the first-order phase transition, which occurs at $\beta \approx 1.0075$, consistent with earlier Monte Carlo studies that locate the transition near $\beta \simeq 1.01$~\cite{Arnold:2000hf, Bonati:2013ota, Bhanot:1981zg, Lautrup:1980xr}. The estimator reliably tracks the input temperature across the regime, even in the presence of a phase transition.

\section{Thermodynamic Diagnostics}

We conclude by summarizing the numerical results from the four-dimensional compact U(1) lattice gauge theory and by highlighting how the configurational temperature estimator reflects and complements the behavior observed in standard Monte Carlo diagnostics.

\begin{itemize}

\item \textit{Thermalization.}  
The configurational temperature estimator provides a sensitive probe of the thermalization process. During the early stages of the Monte Carlo evolution, conventional observables such as the plaquette, and the estimator $\beta_M$, exhibit similar transient behavior before settling into stationary distributions. As illustrated in Fig.~\ref{fig:therm}, even with limited statistics at early Monte Carlo times, $\beta_M$ closely tracks the system’s approach to equilibrium. While different observables may thermalize on different timescales, the comparatively rapid stabilization of $\beta_M$ makes it a useful diagnostic for monitoring the approach toward thermalization and consistency with the target simulation parameters, rather than as a definitive indicator of full equilibrium.

\begin{figure}[ht]
\centering
\includegraphics[width=8cm]{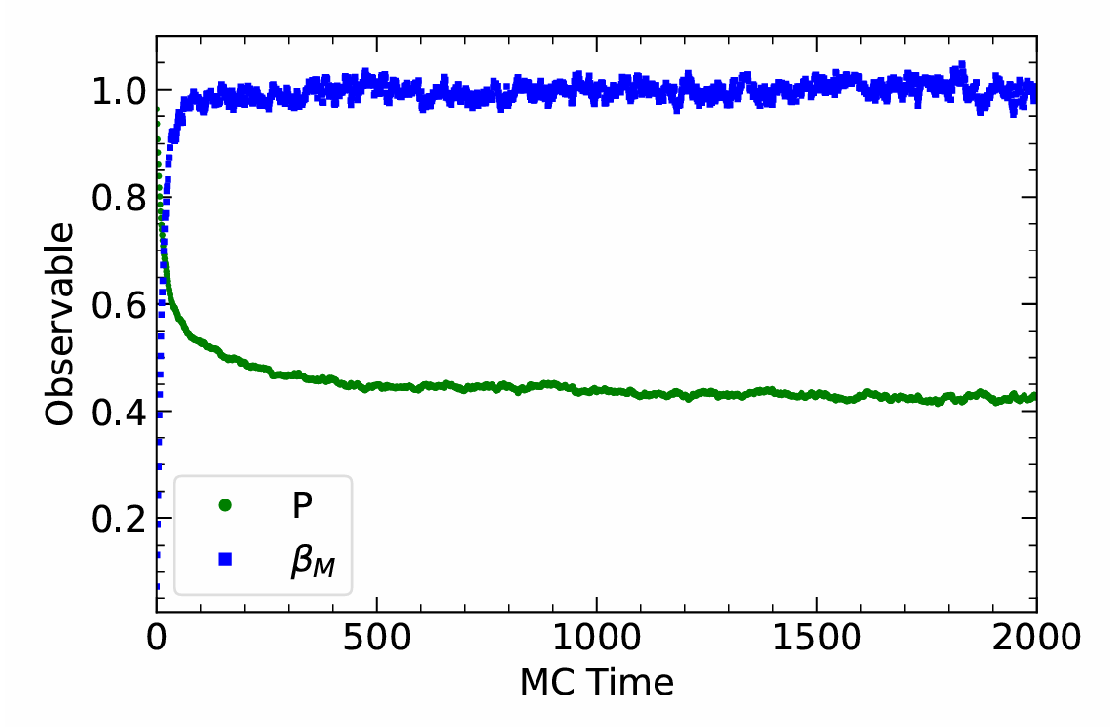}
\caption{The initial Monte Carlo history of the plaquette and $\beta_M$ on an $8^4$ lattice at $\beta = 1.0$, illustrating the thermalization phase.}
\label{fig:therm}
\end{figure}

\item \textit{Error diagnosis.}  
Beyond monitoring thermalization, the estimator is also highly sensitive to algorithmic errors. To demonstrate this, we intentionally introduce an error in the Metropolis update by sampling the acceptance probability from an incorrect uniform interval, $[0.5,1.5]$, rather than the correct range $[0,1]$. This flawed procedure distorts the acceptance rate, alters the equilibrium distribution, and leads to sampling from an incorrect equilibrium distribution. As a result, the measured value of $\beta_M$ deviates systematically from the input $\beta$. Figure~\ref{fig:check} shows a clear separation between the correctly and incorrectly implemented simulations, highlighting the estimator’s effectiveness as a diagnostic tool for detecting subtle implementation errors.

\begin{figure}[ht]
\centering
\includegraphics[width=8cm]{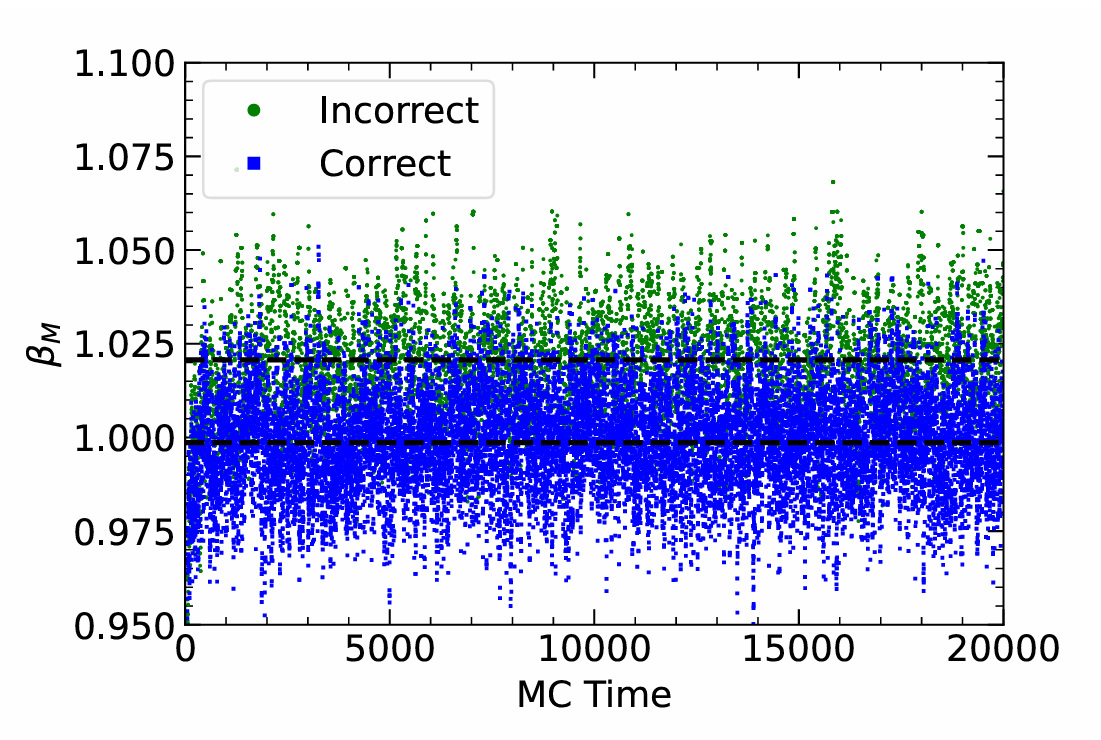}
\caption{Comparison of $\beta_M$ extracted from simulations using correct and incorrect Metropolis updates at $\beta = 1.0$ on an $8^4$ lattice. The estimator clearly identifies the faulty implementation.}
\label{fig:check}
\end{figure}

\item \textit{Phase transition order parameter?}  
Although $\beta_M$ performs well as a diagnostic observable, it is not suitable as an order parameter for identifying phase transitions. At the critical coupling $\beta = 1.0075$, the Monte Carlo histories reveal a marked contrast between the plaquette and the configurational temperature. The plaquette exhibits strong metastability and phase coexistence characteristic of a first-order transition, while $\beta_M$ varies smoothly and shows no clear separation between phases. This behavior, illustrated in Fig.~\ref{fig:order}, indicates that $\beta_M$ does not respond to phase coexistence and should not be interpreted as signaling critical behavior.

\begin{figure}[ht]
\centering
\includegraphics[width=8cm]{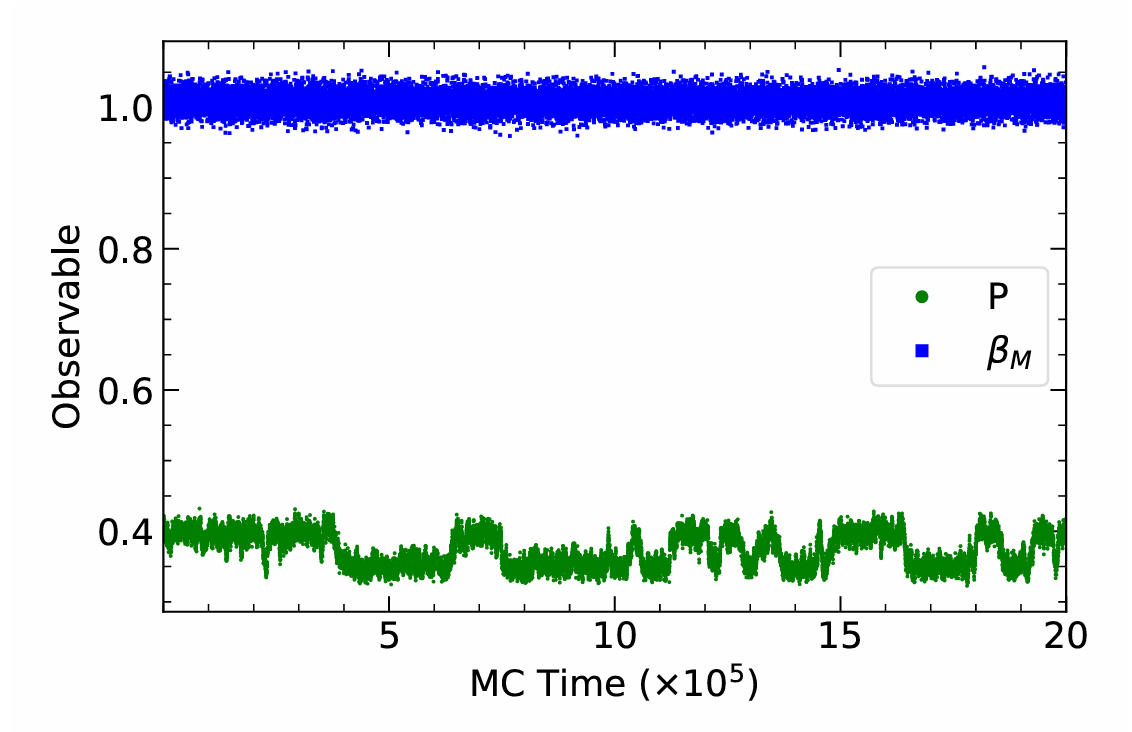}
\caption{Monte Carlo time histories of the plaquette and $\beta_M$ at $\beta = 1.0075$ on an $8^4$ lattice, illustrating the absence of phase separation in $\beta_M$.}
\label{fig:order}
\end{figure}

\end{itemize}

From a computational perspective, the configurational temperature estimator involves evaluating the divergence of a normalized force, which requires both gradient and Hessian information about the lattice action. In practice, this does not necessitate constructing the full Hessian matrix. For local lattice actions, the estimator can be implemented using local second-derivative contributions and directional derivatives along the force, leading to a computational cost comparable to a small number of additional force evaluations per measurement. Since $\beta_M$ is typically measured intermittently during the Monte Carlo evolution, its overall overhead remains modest. The scaling with lattice volume is linear, similar to that of standard force computations, and no qualitatively new bottlenecks are introduced beyond those already present in Hybrid Monte Carlo or Metropolis algorithms. Further details and a more extensive discussion on this can be found in Ref.~\cite{Dhindsa:2025xfv}. Also, see Refs. \cite{Joseph:2025fcd, Joseph:2025xbn} for the application of the configurational temperature estimator as a reliability criterion in complex Langevin simulations.

\section{Summary and Outlook}

In summary, building on Rugh’s geometric formulation of microcanonical temperature, we have constructed a gauge-invariant, purely configurational temperature estimator for lattice gauge theories. We have shown that systematic deviations of the estimator provide a sensitive diagnostic for algorithmic errors in equilibrium Monte Carlo simulations. By incorporating both gradient and Hessian information, the estimator captures essential features of the underlying phase-space geometry. Our results highlight its utility as an important tool for the overall reliability of numerical simulation, without requiring access to dynamical or momentum degrees of freedom.

The extension of the configurational temperature estimator to non-Abelian lattice gauge theories is conceptually straightforward but technically nontrivial. The principal challenge arises from the curved geometry of the gauge group manifold, which necessitates defining gradients and second derivatives in a manner consistent with the group structure and ensuring proper projection onto the Lie algebra. In particular, care must be taken to formulate the divergence of the normalized force so as to preserve gauge invariance and respect the Haar measure. These considerations closely parallel those encountered in force computations for Hybrid Monte Carlo algorithms, suggesting that the estimator can be naturally incorporated into existing non-Abelian simulation frameworks. A detailed implementation and performance analysis for SU($N$) gauge theories is left for future work.

\acknowledgments
The work of A.J. was supported in part by a Start-up Research Grant from the University of the Witwatersrand. 
A.J. gratefully acknowledges the warm hospitality of the National Institute for Theoretical and Computational Sciences (NITheCS) and Stellenbosch University during the NITheCS Focus Area Workshop, {\it Decoding the Universe: Quantum Gravity and Quantum Fields.} 
V.L. acknowledges the support and computational facilities provided by the Indian Institute of Science Education and Research Mohali.

\bibliographystyle{JHEP}
\bibliography{main}

\providecommand{\href}[2]{#2}\begingroup\raggedright\begin{thebibliography}{10}

\bibitem{Rugh:1997}
H.H.~Rugh, \emph{Dynamical approach to temperature}, \href{https://doi.org/10.1103/PhysRevLett.78.772}{\emph{Phys. Rev. Lett.} {\bfseries 78} (1997) 772} [\href{https://arxiv.org/abs/chao-dyn/9701026}{{\ttfamily chao-dyn/9701026}}].

\bibitem{Rugh:1998}
H.H.~{Rugh}, \emph{{A geometric, dynamical approach to thermodynamics}}, \href{https://doi.org/10.1088/0305-4470/31/38/011}{\emph{Journal of Physics A Mathematical General} {\bfseries 31} (1998) 7761} [\href{https://arxiv.org/abs/chao-dyn/9703013}{{\ttfamily chao-dyn/9703013}}].

\bibitem{Dhindsa:2025xfv}
N.S.~Dhindsa, A.~Joseph and V.~Longia, \emph{{Gradient and Hessian-Based temperature estimator in lattice gauge theories: a diagnostic tool for stability and consistency in numerical simulations}}, \href{https://doi.org/10.1007/JHEP10(2025)015}{\emph{JHEP} {\bfseries 10} (2025) 015} [\href{https://arxiv.org/abs/2508.05595}{{\ttfamily 2508.05595}}].

\bibitem{Caselle:2025bgu}
M.~Caselle, A.~Mariani, M.~Panero and A.~Smecca, \emph{{On the equation of state of U(1) lattice gauge theory in three dimensions}}, \href{https://doi.org/10.1007/JHEP03(2025)130}{\emph{JHEP} {\bfseries 03} (2025) 130} [\href{https://arxiv.org/abs/2501.16185}{{\ttfamily 2501.16185}}].

\bibitem{Vettorazzo:2004cr}
M.~Vettorazzo and P.~de~Forcrand, \emph{{Finite temperature phase transition in the 4-d compact U(1) lattice gauge theory}}, \href{https://doi.org/10.1016/j.physletb.2004.10.042}{\emph{Phys. Lett. B} {\bfseries 604} (2004) 82} [\href{https://arxiv.org/abs/hep-lat/0409135}{{\ttfamily hep-lat/0409135}}].

\bibitem{Arnold:2000hf}
G.~Arnold, T.~Lippert, K.~Schilling and T.~Neuhaus, \emph{{Finite size scaling analysis of compact QED}}, \href{https://doi.org/10.1016/S0920-5632(01)01001-5}{\emph{Nucl. Phys. B Proc. Suppl.} {\bfseries 94} (2001) 651} [\href{https://arxiv.org/abs/hep-lat/0011058}{{\ttfamily hep-lat/0011058}}].

\bibitem{Bonati:2013ota}
C.~Bonati and M.~D'Elia, \emph{{Phase diagram of the 4D U(1) model at finite temperature}}, \href{https://doi.org/10.1103/PhysRevD.88.065025}{\emph{Phys. Rev. D} {\bfseries 88} (2013) 065025} [\href{https://arxiv.org/abs/1305.3564}{{\ttfamily 1305.3564}}].

\bibitem{Bhanot:1981zg}
G.~Bhanot, \emph{{The Nature of the Phase Transition in Compact {QED}}}, \href{https://doi.org/10.1103/PhysRevD.24.461}{\emph{Phys. Rev. D} {\bfseries 24} (1981) 461}.

\bibitem{Lautrup:1980xr}
B.E.~Lautrup and M.~Nauenberg, \emph{{Phase Transition in Four-Dimensional Compact QED}}, \href{https://doi.org/10.1016/0370-2693(80)90400-1}{\emph{Phys. Lett. B} {\bfseries 95} (1980) 63}.

\bibitem{Joseph:2025fcd}
A.~Joseph and A.~Kumar, \emph{{Thermodynamic Diagnostics for Complex Langevin Simulations: The Role of Configurational Temperature}},  \href{https://arxiv.org/abs/2509.08287}{{\ttfamily 2509.08287}}.

\bibitem{Joseph:2025xbn}
A.~Joseph and A.~Kumar, \emph{{Configurational Temperature as a Diagnostic for Complex Langevin Dynamics in the 3D XY Model}},  \href{https://arxiv.org/abs/2509.13314}{{\ttfamily 2509.13314}}.

\end{thebibliography}\endgroup
\end{document}